\documentclass[aps,prb,twocolumn,showpacs,showkeys,floatfix]{revtex4}

\usepackage{graphicx}
\graphicspath{{figs/}}
\begin{document}
\title{Graphene-based torsional resonator from molecular dynamics simulation}
\author{Jin-Wu~Jiang}
    \altaffiliation{Electronic address: phyjj@nus.edu.sg}
    \affiliation{Department of Physics and Centre for Computational Science and Engineering,
             National University of Singapore, Singapore 117542, Republic of Singapore }
\author{Jian-Sheng~Wang}
    \affiliation{Department of Physics and Centre for Computational Science and Engineering,
                 National University of Singapore, Singapore 117542, Republic of Singapore }

\date{\today}
\begin{abstract}
Molecular dynamics simulations are performed to study graphene-based torsional mechanical resonators. The quality factor is calculated by $Q_{F}=\omega\tau/2\pi$, where the frequency $\omega$ and life time $\tau$ are obtained from the correlation function of the normal mode coordinate. Our simulations reveal the radius-dependence of the quality factor as $Q_{F}=2628/(22R^{-1}+0.004R^{2})$, which yields a maximum value at some proper radius $R$. This maximum point is due to the strong boundary effect in the torsional resonator, as disclosed by the temperature distribution in the resonator. Resulting from the same boundary effect, the quality factor shows a power law temperature-dependence with power factors bellow 1.0. The theoretical results supply some valuable information for the manipulation of the quality factor in future experimental devices based on the torsional mechanical resonator.
\end{abstract}

\maketitle

\pagebreak

\section{introduction}
Mechanical resonators have practical applications in fields like accurate mass or force sensor.\cite{SazonovaV,MeyerJC} The mechanical resonator is built based on a special natural vibration of the system. A gigahertz resonator was proposed theoretically using multi-walled carbon nanotubes,\cite{ZhengQS2002} which was inspired by a retraction effect in this system.\cite{CumingsJ} This resonator is built on the oscillating vibration of multi-walled carbon nanotubes. The self-retracting effect, which is similar as resonator, was also observed in multiple graphene layers.\cite{ZhengQS2008} Using the cantilevered carbon nanotube, several experimental groups have constructed a beam resonator, which actually takes advantage of the bending vibration (i.e. flexure mode).\cite{SazonovaV,JiangH,WitkampB} Besides the bending vibration, the torsional vibration of the single-walled carbon nanotube has also been used to create a torsional pendulum to measure the torsional force precisely.\cite{MeyerJC} For graphene, both experimental\cite{BunchJS,SanchezG,Zande} and theoretical works\cite{AtalayaJ,KimSY,KimSY2} have so far focused on the study of resonators based on the bending vibration. However, there is still no study on possible mechanical resonator based on the torsional vibration of graphene.

In this paper, we apply molecular dynamics (MD) simulation to record the correlation function of the normal mode coordinate. The frequency $\omega$ and life time $\tau$ of the torsional vibration are obtained from the correlation function. We then calculate the quality factor ($Q_{F}$) of the torsional resonator by $Q_{F}=\omega\tau/2\pi$. Calculations focus on the size dependence and temperature dependence of the quality factor, which show the importance of the boundary effect. We also address an interesting recurrent effect in finite sized resonators and estimate the corresponding recurrent speed.

\begin{figure}[htpb]
  \begin{center}
    \scalebox{1.0}[1.0]{\includegraphics[width=8cm]{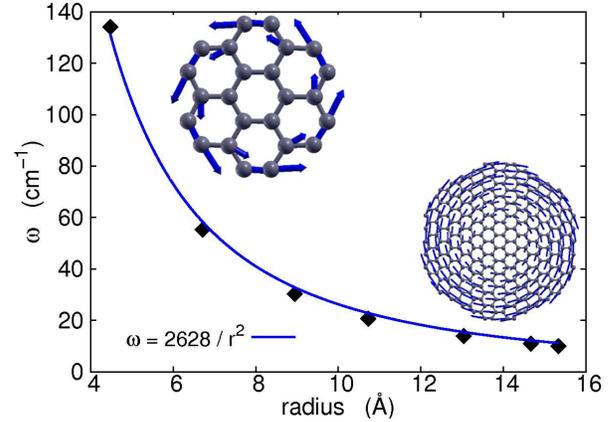}}
  \end{center}
  \caption{(Color online) Frequency versus the radius of the graphene torsional resonator. Insets are two resonator samples with radius 4.5~{\AA} and 15.5~{\AA}.}
  \label{fig_frequency}
\end{figure}
\section{computational method}
The structure of two resonators are shown as insets of Fig.~\ref{fig_frequency}. The central six carbon atoms are fixed during the simulation. Free boundary condition is applied for carbon atoms at the edge of the resonator. The arrow attached to each atom exhibits its vibrational amplitude in the torsional vibration. The maximum vibrational amplitudes are at the edge of the resonator. As a result, there is no stress at the edge, which is required by the free boundary condition. The torsional resonator can be actuated according to this vibrational morphology. The interaction between carbon atoms is described by the Brenner potential in the MD simulation.\cite{Brenner} The No\'se Hoover heat bath\cite{Nose,Hoover} is applied to realize constant temperature when necessary. Newton equations are solved in terms of the velocity Verlet algorithm with a time step of 1 fs.

A standard MD simulation for resonators is realized in three stages.\cite{AtalayaJ,KimSY} In stage one, the resonator system is thermalized to a constant temperature with the help of a heat bath. In stage two, the heat bath is removed and the mechanical resonator is actuated by generating the corresponding natural vibration of the system. The phonon vibration is typically excited by displacing each atom according to the eigen vector of the vibration. In the final stage, the resonator system is allowed to oscillate freely, where the quality factor is calculated from the energy damping of the oscillation. We are not going to follow this standard procedure. Instead, we plan to perform MD simulation for the resonator in two steps. In step one, the system is thermalized at a constant temperature, which is realized by the No\'se Hoover heat bath.\cite{Nose,Hoover} This step is the same as the first stage in the standard procedure. In step two, the heat bath is removed and the resonator is simulated in a microcanonical ensemble. In this step, we calculate the natural vibrations of the resonator system by solving an eigen value problem, and we can get the eigen vectors $\xi^{k}$ for each phonon mode $k$. We then find the eigen vector corresponding to the torsional vibration ($\xi^{torsion}$). Using this eigen vector, $\xi^{torsion}$, we can obtain the normal mode coordinate $Q(t)$ of the torsional mode by following projection technique:
\begin{eqnarray}
Q(t)=\frac{1}{\sqrt{N}}\sum_{j=1}^{3N} \xi^{torsion}_{j} (r_{j}-r^{0}_{j})\nonumber,
\label{eq_Q}
\end{eqnarray}
where $3N$ is the total degree of freedom. $(r_{j}(t)-r^{0}_{j})$ is the displacement of the j-th degree of freedom in the MD simulation. $r_{j}(t)$ is the position of the j-th degree of freedom at time $t$ and $r_{j}^{0}$ is the equilibrium position. The time dependence of the normal mode coordinate, $Q(t)$, is determined by the trajectory history from MD simulation. The quality factor of the torsional mechanical resonator is calculated by the definition $Q_{F}=\nu \tau=\omega \tau/2\pi$.\cite{JorioA} The frequency $\omega$ and life time $\tau$ of the torsional vibration can be derived from the correlation function. Under the single mode relaxation time approximation, the correlation function can be calculated by:\cite{ZwanzigR}
\begin{eqnarray}
\langle Q^{\sigma}(t)Q^{\sigma}(0) \rangle/\langle Q^{\sigma}(0)Q^{\sigma}(0) \rangle=\cos (\omega^{\sigma} t) e^{-t/\tau^{\sigma}},
\label{eq_QQ}
\end{eqnarray}
where $Q^{\sigma}(t)$ is the normal mode coordinate of the phonon mode $\sigma$. $Q^{\sigma}(0)$ is the value at the beginning of the second step as described in above. $\omega^{\sigma}$ is the frequency and $\tau^{\sigma}$ is the life time of this mode. There are two different interpretations for the average in the correlation function. It can be realized by the average over MD simulation time:
\begin{eqnarray}
\langle Q^{\sigma}(t_{n})Q^{\sigma}(0) \rangle = \frac{1}{M}\sum_{m=1}^{M}Q^{\sigma}(t_{n}+t_{m})Q^{\sigma}(t_{m}),
\label{eq_averageovertime}
\end{eqnarray}
where integer $M$ is the total MD simulation step. $M$ is usually in an order of $10^{7}$, so that the standard deviation can be small enough.
\begin{figure}[htpb]
  \begin{center}
    \scalebox{1.0}[1.0]{\includegraphics[width=8cm]{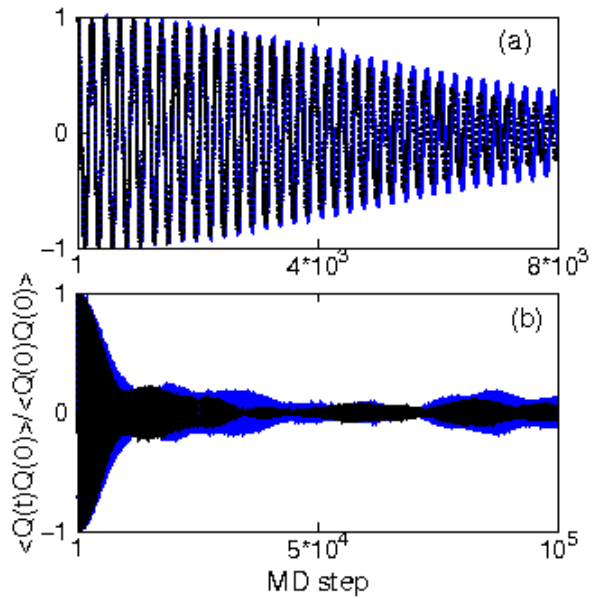}}
  \end{center}
  \caption{(Color online) Correlation function of the torsional normal mode coordinate from averaging over simulation time. The blue curve is for a total simulation time of $10^{6}$ simulation steps, while the black curve corresponds to a total simulation time of $10^{7}$ simulation steps. Panel (b) shows the correlation function in a larger time scale.}
  \label{fig_QQ1}
\end{figure}
 A long simulation time is required to reduce the statistical error due to the small size of the systems in our studied. Correlation functions for resonator with radius 4.5~{\AA} from this averaging with $M=10^{6}$ (in blue) and $10^{7}$ (in black) are shown in Fig.~\ref{fig_QQ1}. Panel (b) is for a larger time scale. This figure displays that the results show some dependence on the total simulation time $M$, which may be due to possible statistical errors for large $M$. An interesting recurrent effect can be observed from panel (b). The correlation function does not decrease monotonously. Instead, it arises at some particular time $t_{r}$ after its first decaying. The recurrence is due to finite size effect and is not able to be eliminated by increasing simulation time.\cite{ZwanzigR} The physical picture of the recurrence is that the torsional phonon travels along the edge of the circular resonator and will come back to its starting point after some time. We will show further discussions on the recurrent effect in next section.

\begin{figure}[htpb]
  \begin{center}
    \scalebox{1.0}[1.0]{\includegraphics[width=8cm]{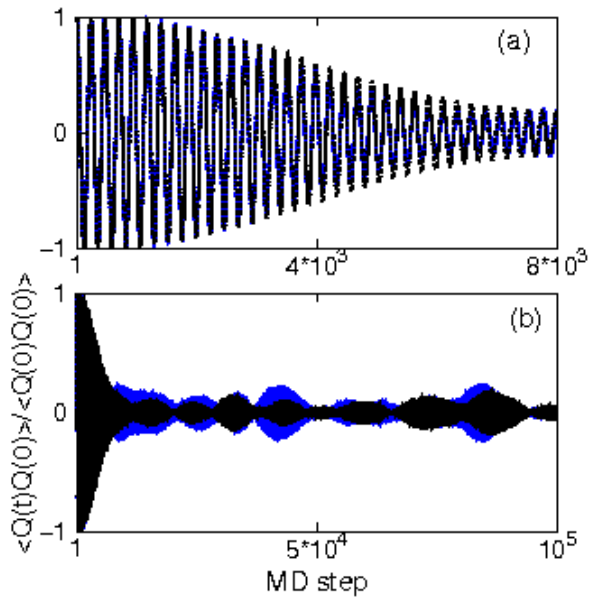}}
  \end{center}
  \caption{(Color online) Correlation function of the torsional normal mode coordinate from averaging over initial condition of the molecular dynamics simulation. The blue curve is the average over 300 different initial conditions, while \textbf{the} black curve corresponds to an average over 450 different initial conditions. Panel (b) shows the correlation function in a larger time scale.}
  \label{fig_QQ2}
\end{figure}
Another interpretation for the average in the correlation function of Eq.~(\ref{eq_QQ}) is to average over initial conditions:
\begin{eqnarray}
\langle Q^{\sigma}(t_{n})Q^{\sigma}(0) \rangle = \frac{1}{N}\sum_{i=1}^{N}Q^{\sigma}_{i}(t_{n})Q^{\sigma}_{i}(0),
\label{eq_averageoverinitial}
\end{eqnarray}
where $i=1,2,3,...,N$ is the steady state from different initial conditions. This averaging method can help to avoid possible statistical error induced by extremely long simulation time. We will use this averaging approach in following. Correlation functions in resonator with radius 4.5~{\AA} from this averaging method with $N=300$ (in blue) and 450 (in black) are shown in Fig.~\ref{fig_QQ2}. Panel (b) is for a larger time scale. The recurrent effect also exists in this averaging method as shown in panel (b). The frequency and life time are obtained by fitting Eq.~(\ref{eq_QQ}) to the correlation functions from MD simulation in Fig.~\ref{fig_QQ2}~(a). The recurrent effect can be partially avoided by finishing the fitting procedure before the recurrence happens. That is why panel (a) is used in the fitting procedure rather than panel (b).
\begin{figure}[htpb]
  \begin{center}
    \scalebox{1.0}[1.0]{\includegraphics[width=8cm]{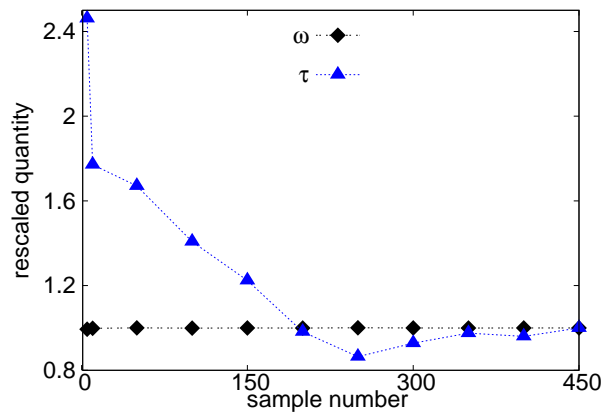}}
  \end{center}
  \caption{(Color online) The value of the frequency ($\omega$) and life time ($\tau$) of the torsional resonator with radius 4.5~{\AA} versus the sampling number of the initial conditions. The displayed data is the original value rescaled by the value at sampling number 450.}
  \label{fig_samplenumber}
\end{figure}
 In the following, the fitting procedure is done for the decay of the correlation function before recurrent time $t_{r}$. Fig.~\ref{fig_samplenumber} shows the frequency and life time for resonator with radius 4.5~{\AA} from different sampling number $N$. The value of these two quantities in the figure have been rescaled by the value at $N=450$. The frequency has almost the same value in the whole range of $N$. The life time reaches a saturated value after $N>300$. We always use $N=350$ for following calculations.

\section{results and discussion}
Fig.~\ref{fig_frequency} shows the natural frequency at 0 K for resonators of different radii. The radius dependence of the frequency can be well described by function $\omega=2628/R^{2}$, where $R$ is the radius of the resonator in \AA. This behavior is similar as a circular elastic solid thin plate under the same boundary condition. Treating the graphene resonator as a circular elastic thin plate, we can obtain its torsional vibrational frequency: $\omega=\frac{1}{R^{2}}\sqrt{\frac{k_{\theta}S_{0}}{6\pi m_{n}}}$,\cite{ClelandAN} where $m_{n}$ is the mass of neutron. $S_{0}=2.62\AA^{2}$ is the area of the two-atom unit cell in graphene sheet.\cite{SaitoR} The effective torsional constant $k_{\theta}$ is estimated to be about 182.55 eV from the data in Fig.~\ref{fig_frequency}. This value is larger than the Young's modules of the graphene which is about 50 eV,\cite{PopovVN,JiangJW,JiangJW2009} indicating stronger stability for resonator under torsional strain than the uniaxial strain. \textbf{In usual, higher operational (resonant) frequency is favorable for the sensitivity of nanoscale resonators. A sensitive mass sensor with resolution $10^{-18}$ g can be obtained from carbon-nanotube-based electromechanical resonators with the operational frequency over 1.3 GHz.\cite{PengHB} Fig.~\ref{fig_frequency} shows that the resonant frequency is over 1.0 GHz for torsional resonators with radius smaller than 70 nm. This is helpful for the application of torsional resonators as sensitive mass sensor.}

\begin{figure}[htpb]
  \begin{center}
    \scalebox{1.0}[1.0]{\includegraphics[width=8cm]{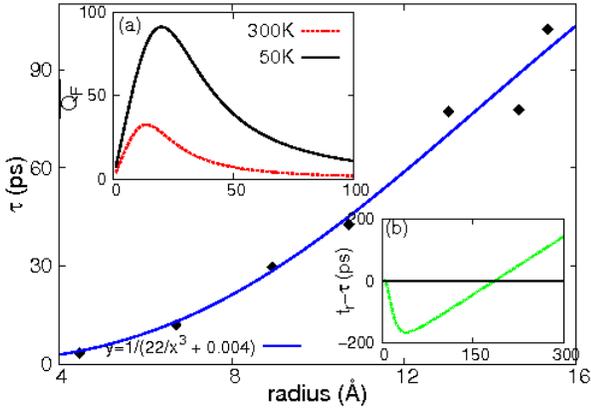}}
  \end{center}
  \caption{(Color online) The room temperature life time versus the radius of the torsional resonator. Inset (a) is the quality factor from $Q_{F}=\omega\tau/2\pi$, with $\omega$ from the formula in Fig.~\ref{fig_frequency} and life time by the fitting formula. Inset (b) is the  difference between the recurrent time $t_{r}=2\pi R/0.5$ and the life time $\tau$ at room temperature. The x axis of both insets are the same as the main figure.}
  \label{fig_radius}
\end{figure}

Fig.~\ref{fig_radius} displays the life time at room temperature for the seven resonators with radius varying from 4.5~{\AA} to 16~{\AA}. These points fall in a curve described by the function $\tau=1/(22R^{-3}+0.004)$ ps. The life time increases with increasing radius,
\begin{figure}[htpb]
  \begin{center}
    \scalebox{1.0}[1.0]{\includegraphics[width=8cm]{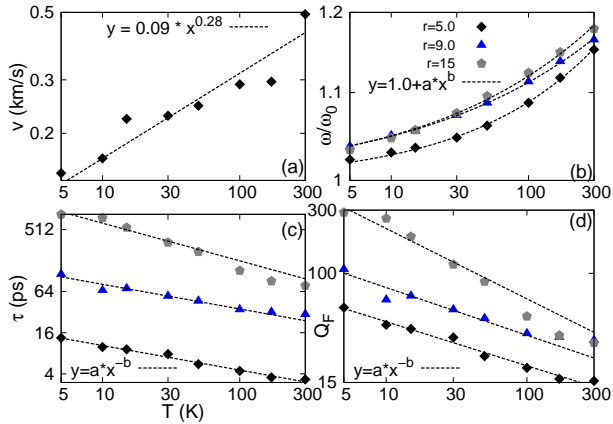}}
  \end{center}
  \caption{(Color online) Temperature dependence for different physical quantities in the torsional resonator. (a). Recurrent velocity in resonator with radius 4.5~{\AA}. (b). The frequency in resonators with three different radius. (c). The life time of the torsional vibration mode in these three resonators. (d). The quality factor of these three torsional resonators.}
  \label{fig_temperature}
\end{figure}
 which implies strong boundary scattering effect in the resonator. The boundary effect is of particular importance for free edges.\cite{KimSY} It is not easy to perform MD simulation for very large resonators, yet the fitting expression for the life time can be combined with the frequency $\omega=2628/R^{2}$ to predict the quality factor of some large resonators at room temperature. This prediction is shown as inset (a) in the figure. The frequency formula at 0 K has been used to predict the room temperature quality factor, since $\omega$ is not sensitive to temperature. The quality factor at 50 K has also been obtained by similar procedure. Both curves in inset (a) display a maximum value for the quality factor at some proper radius. For smaller radius, the boundary scattering is more important than the phonon-phonon scattering, so the increase of life time overpasses the decrease of frequency, leading to the increase of quality factor. In large resonator, the phonon-phonon scattering is more important, so the situation becomes opposite and the quality factor decreases with increasing radius. For a given radius, the phonon-phonon scattering is stronger at higher temperature. Hence, the phonon-phonon scattering will become more important than the boundary scattering in resonators with smaller radii and at higher temperature. As a result, the maximum quality factor is achieved at smaller radius if temperature is higher.

\begin{figure}[htpb]
  \begin{center}
    \scalebox{1.0}[1.0]{\includegraphics[width=8cm]{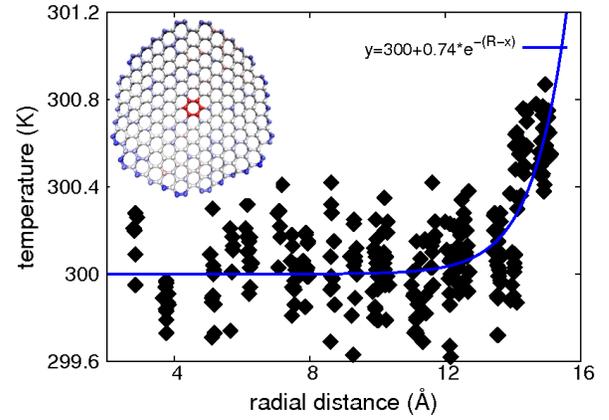}}
  \end{center}
  \caption{(Color online) The temperature distribution in the torsional resonator with radius 15.5~{\AA}. Inset shows the spatial temperature distribution by color.}
  \label{fig_temprofile}
\end{figure}
With the recurrent time $t_{r}$ and the circumference of resonator $L=2\pi R$, we can estimate the recurrent velocity $v=L/t_{r}$. Fig.~\ref{fig_temperature}~(a) shows the temperature dependence for the recurrent velocity of the resonator with $R=4.5$~{\AA}. The recurrent velocity is larger at higher temperature, which is a common phenomenon in the nature. The curve can be fitted to a power function $v=0.09T^{0.28}$ km/s. $v\approx 0.5$ km/s at room temperature. This value is close to the sound speed in the air, while it is much smaller than the longitudinal and transverse acoustic velocities in the graphene,\cite{SaitoR} which is probably related to the edge effect in the torsional graphene resonator.

We are interesting to investigate how the recurrence affects the correlation function, although we have only used the function for $t<t_{r}$. In case of $t_{r}<\tau$, the recurrent effect happens before the damping of the torsional vibration, so the recurrent wave will couple to the original wave in the correlation. This effect causes some unavoidable error for the correlation function after $t>t_{r}$. That's another reason why we finish the fitting process before $t<t_{r}$. The inset (b) in Fig.~\ref{fig_radius} shows the curve of $t_{r}-\tau$ at room temperature, where $t_{r}=2\pi R/v$ with constant $v=0.5$ km/s as the recurrent velocity at 300 K. It shows that the recurrent effect will have influence for resonator with $R<180$~{\AA}. If the MD simulation can be done for resonator with radius $R>180$~{\AA}, the recurrent effect will be much weaker on the fitting procedure of the correlation function. For extremely large resonator, there will be no recurrent effect as $t_{r}$ is almost infinity.

Fig.~\ref{fig_temperature}~(b) shows the temperature dependence for the frequency of three resonators with radius $R=$ 4.5, 9.0, and 15.0~{\AA}. For all three resonators, a blue-shift of the frequency is observed with the increase of temperature. These three curves can be fitted to functions: $\omega/\omega_{0}=1+a*T^{b}$, with parameters $(a,b)$ as $(0.01, 0.5)$, $(0.02, 0.36)$, and $(0.02, 0.38)$. Overall, the frequency is not sensitive to temperature and the shift is within $20\%$ below 300 K. We expect this blue-shift can be measured in the experiment, while current experiments focus on the temperature dependence of other vibration modes such as radial breathing modes in the carbon nanotube.\cite{Raravikar} Panel (c) is the life time at different temperatures. They can be described by power functions $\tau=a*T^{-b}$, where parameters $(a, b)$ are $(23.8, 0.36)$, $(186.3,0.36)$, and $(2255.8, 0.55)$ in three resonators. It is interesting that the power factor $b$ is less than 1.0 in all of the three resonators. The phonon life time is inversely proportional to the temperature in bulk materials,\cite{HollandMG} because the three-phonon scattering is the major process for the damping of the phonon vibration in bulk materials. However, in nano-materials such as the graphene resonator, the boundary effect provides another channel for phonon decaying, which is temperature independent. The combination of the phonon-phonon scattering and the boundary effect leads to a power factor $b$ less than 1.0. For larger resonator with $R=15.0$~{\AA}, the power factor $b=0.55$ is closer to the value of 1 than the other two smaller resonators where $b=0.36$. It implies that the boundary effect is weaker in larger resonator, which is quite reasonable. As a result of the temperature dependence of the life time, the quality factor shows similar behavior in panel (d), where the two parameters $(a,b)$ in the power function are (92.7, 0.33), (178.9, 0.36), and (739.7, 0.53). Due to the boundary effect, the quality factor is very low for the torsional resonator simulated in our work. The boundary effect on the quality factor has also been pointed out by Kim and Park.\cite{KimSY} \textbf{If the atoms on the boundary are fixed and let the atoms at the center free, the resonator will undergo a bending oscillation instead of torsional oscillation. In this type of circular bending resonator, Kim and Park have shown that the scattering by free edges will be considerably removed. As a result, the quality factor can be higher than 2000 at room temperature and increase to be $10^{5}$ at 10 K.\cite{KimSY}} In Fig.~\ref{fig_temprofile}, the boundary effect is disclosed by the temperature distribution in the resonator. The temperatures of the inner six carbon atoms are zero, as they are fixed during MD simulation. There is a temperature jump between the edge atoms and the inner carbon atoms, which results from the edge modes localized in this free boundary region.\cite{SanchezG,JiangJW2009b}

We further remark that the torsional resonator described in our manuscript may be difficult to be realized in the experiment currently. The most difficult part is to fix the central region. A possible idea may be to fix the central region by hybridizing it to a silicon nanowire or diamond nanowire. Once the central region is fixed, it is straight forward to generate the torsional resonator by displacing the edge of the resonator using scanning electron microscope, which has been successfully applied to push the surface graphene layer away from the graphite.\cite{ZhengQS2008} Although the torsional resonator is not so easy to be realized with current experimental techniques, it is possible to be produced in the future with the development of the technique and some smart skills from experimentalists. Forgive the detailed technical difficulties in preparing the device, the significant contribution from our work is that we have proposed another theoretical approach to study the quality factor of the nanoresonator. Currently, there is only one MD simulation method for the investigation of mechanical resonators.\cite{JiangH,KimSY} Our method can be used to obtain the pure intrinsic damping mechanism for the nanoresonator, as the quality factor is calculated from an equilibrium MD simulation and no mechanical oscillation is generated.

It should be mentioned that we study the intrinsic damping mechanisms for the resonator, which is due to the phonon phonon scattering. This intrinsic damping mechanism depends on the structure and the interaction of the system. That is why we can calculate the quality factor from the correlation function of the normal mode coordinate. The phonon phonon scattering is described by the Brenner potential,\cite{Brenner} which can successfully capture both linear and nonlinear properties of the carbon-based system.

\section{conclusion}
To conclude, we perform MD simulation to investigate graphene-based torsional resonators. The quality factor is calculated by $Q_{F}=\omega \tau /2\pi$, where $\omega$ and $\tau$ are the frequency and life time of the torsional vibration. $\omega$ and $\tau$ are obtained from the correlation function of the normal mode coordinate. As a result of the boundary effect, the quality factor is small and has a maximum value for resonator with proper radius. The maximum quality factor is achieved with smaller radius at higher temperature resulting from the interplay between boundary effect and the phonon-phonon scattering. The quality factor depends on temperature as a power law function $Q_{F}=a*T^{-b}$ with the power factor $b<1$, which is due to the same boundary effect. We also investigate the recurrent effect in finite sized mechanical resonators and the recurrent speed is found to be around 0.5 km/s at room temperature.

\textbf{Acknowledgements} The authors thank Dr X. F. Xu for helpful discussions and Prof. H. S. Park at Boston University for valuable communication. The work is supported by a URC grant of R-144-000-257-112 of National University of Singapore.


\begin{thebibliography}{}
\bibitem{SazonovaV} V. Sazonova, Y. Yaish, H. Ustunel, D. Roundy ,T. A. Arias, and P. L. McEuen, Nature \textbf{431}, 284 (2004).

\bibitem{MeyerJC} J. C. Meyer, M. Paillet, and S. Roth, Science \textbf{309}, 1539 (2005).

\bibitem{ZhengQS2002} Q. S. Zheng and Q. Jiang, Phys. Rev. Lett. \textbf{88}, 045503 (2002).

\bibitem{CumingsJ} J. Cumings and A. Zettl, Science \textbf{289}, 602 (2000).

\bibitem{ZhengQS2008} Q. S. Zheng, B. Jiang, S. Liu, Y. Weng, L. Lu, Q. Xue, J. Zhu, Q. Jiang, S. Wang, and L. Peng, Phys. Rev. Lett. \textbf{100}, 067205 (2008).

\bibitem{JiangH} H. Jiang, M.-F. Yu, B. Liu, and Y. Huang, Phys. Rev. Lett. \textbf{93}, 185501 (2004).

\bibitem{WitkampB} B. Witkamp, M. Poot, and H. S. J. van der Zant, Nano. Lett. \textbf{6}, 2904 (2006).

\bibitem{BunchJS} J. S. Bunch, A. M. van der Zande, S. S. Verbridge, I. W. Frank, D. M. Tanenbaum, J. M. Parpia, H. G. Craighead, and P. L. McEuen, Science \textbf{315}, 490 (2007).

\bibitem{SanchezG} G. Sanchez, A. M. van der Zande, A. San Paulo, B. Lassagne, P. L. McEuen, and A. Bachtold, Nano. Lett. \textbf{8}, 1399 (2008).

\bibitem{Zande} A. M. van der Zande, R. A. Barton, J. S. Alden, C. S. Ruiz-Vargas, W. S. Whitney, P. H. Q. Pham, J. Park, Jeevak M. Parpia, H. G. Craighead, and P. L. McEuen, Nano. Lett. \textbf{10}, 4869 (2010).

\bibitem{AtalayaJ} J. Atalaya, A. Isacsson, and J. M. Kinaret, Nano. Lett. \textbf{8}, 4196 (2008).

\bibitem{KimSY} S. Y. Kim and H. S. Park, Nano. Lett. \textbf{9}, 969 (2009).

\bibitem{KimSY2} S. Y. Kim and H. S. Park, Appl. Phys. Lett. \textbf{94}, 101918 (2009).

\bibitem{Brenner} D. W. Brenner, O. A. Shenderova, J. A. Harrison, S. J. Stuart, B. Ni and S. B. Sinnott, J. Phys.:Condens. Matter \textbf{14}, 783 (2002).

\bibitem{Nose} S. No\'se, J. Chem. Phys. \textbf{81}, 511 (1984).

\bibitem{Hoover} W. G. Hoover, Phys. Rev. A, \textbf{31}, 1695 (1985).

\bibitem{JorioA} A. Jorio, G. Dresselhaus, and M. S. Dresselhaus, \textit{Topics in Applied Physics (Volume III) Carbon Nanotubes: Advanced Topics in the Synthesis, Structure, Properties} (Springer-Verlag, Berlin Heidelberg, 2008) page 381.

\bibitem{ZwanzigR} R. Zwanzig, \textit{Nonequilibrium Statistical Mechanics} (Oxford University press, New York, 2001).

\bibitem{ClelandAN} A. N. Cleland, \textit{Foundations of Nanomechanics} (Springer-Verlag, Berlin Heidelberg, 2003).

\bibitem{SaitoR} R. Saito, G. Dresselhaus, and M. S. Dresselhaus, \textit{Physical Properties of Carbon Nanotubes} (Imperial College Press, London, 1998).

\bibitem{PopovVN} V. N. Popov, V. E. Van Doren, and M. Balkanski, Phys. Rev. B \textbf{61}, 3078 (2000).

\bibitem{JiangJW} J.-W. Jiang, H. Tang, B.-S. Wang and Z.-B. Su, J. Phys.: Condens. Matter \textbf{20}, 045228 (2008).

\bibitem{JiangJW2009} J.-W. Jiang, J.-S. Wang, and B. Li, Phys. Rev. B \textbf{80}, 113405 (2009).

\bibitem{PengHB} H. B. Peng, C. W. Chang, S. Aloni, T. D. Yuzvinsky, and A. Zettl, Phys. Rev. Lett. \textbf{97} 087203 (2006).

\bibitem{Raravikar} N. R. Raravikar, P. Keblinski, A. M. Rao, M. S. Dresselhaus, L. S. Schadler, and P. M. Ajayan, Phys. Rev. B \textbf{66}, 235424 (2002).

\bibitem{HollandMG} M. G. Holland, Phys. Rev. \textbf{132}, 2461 (1963).

\bibitem{JiangJW2009b} J. W. Jiang, J. Chen, J.-S. Wang, and B. Li, Phys. Rev. B \textbf{80}, 052301 (2009).
\end{thebibliography}
\end{document}